# Complex Dynamic Systems in Education: Beyond the Static, the Linear and the Causal Reductionism

Mohammed Saqr[1], Daryn Dever[2], Sonsoles López-Pernas[1],
Christophe Gernigon[3], Gwen Marchand[4], Avi Kaplan[5]

[1] *University of Eastern Finland,* [2] *University of Florida,* [3] *Université de Montpellier,*
[4] *University of Nevada,* [5] *Temple University*

## Abstract

Traditional methods in educational research often fail to capture the complex and evolving nature of learning processes. This chapter examines the use of complex systems theory in education to address these limitations. The chapter covers the main characteristics of complex systems such as non-linear relationships, emergent properties, and feedback mechanisms to explain how educational phenomena unfold. Some of the main methodological approaches are presented, such as network analysis and recurrence quantification analysis to study relationships and patterns in learning. These have been operationalized by existing education research to study self-regulation, engagement, and academic emotions, among other learning-related constructs. Lastly, the chapter describes data collection methods that are suitable for studying learning processes from a complex systems' perspective.

## Introduction

Learning —like most psychological phenomena— is far from being simple or trivial. It involves multiple cognitive processes, contexts and dimensions with complex and dynamic interplay between them (Hilpert and Marchand 2018; Kaplan and Garner 2020; Saqr et al. 2024d). Therefore, most learning theories and frameworks describe learning as a multifaceted process that evolves over time. For example, most theorists depict self-regulated learning as having several phases or processes (Panadero 2017) that influence each other resulting in the emergence of certain behaviors, e.g., learning strategies. Similarly, engagement is commonly conceptualized as a multidimensional process that unfolds in time with a significant interplay between the dimensions across contexts, tasks and cultures (Saqr et al. 2024d; Symonds et al. 2024). These complexity features apply to most learning theories, frameworks, and constructs and also extends to interpersonal processes, groups, classrooms, and organizations (Gouvea 2023). All of the aforementioned examples, and many others, exhibit features typical of complex systems. Traditional methods often fail to adequately capture their complexity and dynamics, and therefore, complex system methods are required to capture their unique features (Koopmans 2020; Saqr et al. 2024d).

But, before moving on, we must articulate what we mean by a complex system. The word *complexity* signifies multiple components that interact with each other in intricate ways (Thurner et al. 2018). Thus, a *complex system* is a collection of components that interact together in unique ways that lead to the emergent behavior of the system as a whole and cannot be understood by nor reduced to its individual components. The term *dynamic* indicates that the system changes and evolves over time exhibiting different states and conditions (Sayama 2015; Scheffer 2020).

Researchers vary in their definitions of complex dynamic systems and in the level of detail by which they describe these systems. However, most researchers can agree on some basic requirements for identifying a system as complex: the presence of intricate properties, chaotic unpredictable behavior or outcomes, nonlinear dynamics, and interactions between the system components (Olthof et al. 2023; Bringmann et al. 2023). Furthermore, these properties exist in different mixtures exhibiting different dynamics and interactions and due to this variability, complex systems and their associated behaviors/properties are not perfectly replicable (Sayama 2015; Scheffer 2020).

# Why complex systems in education?

As we have mentioned before, humans, their behavior, and their social structures (e.g., groups, institutions, societies) manifest complex and dynamic behavior, making conceptualizing them as complex dynamic systems an apt epistemological endeavor (Hilpert and Marchand 2018). Viewing human phenomena as complex dynamic systems allows us to capture their complex organization, dynamics, and behavior (López-Pernas and Saqr 2024). It stands to reason that while traditional statistics and inference methods may be useful in capturing some characteristics of complex systems or offer a summary of a system's state at a single time, such methods are incompatible and would fail in fully capturing the complexity of these systems (Bringmann et al. 2023).

Traditional statistics were designed to capture what is commonly known as a box-and-arrow structure, where A and B leads to C; what is known as component-dominant structure. In component-dominant structures, the components themselves in their accumulation, are assumed to define the behavior of the system (López-Pernas and Saqr 2024). Therefore, the focus in complex systems shifts from analyzing components in isolation to the whole system and to understanding the interactions between them (Hilpert and Marchand 2018; Saqr et al. 2024d). The effects of interactions—rather than the individual components themselves—become more significant in determining the system's behavior. These interactions can result in emergent behavior that cannot be fully explained by simply summing the effects of individual components (Hilpert and Marchand 2018; Saqr et al. 2024d).

Let us take an example. In a box-and-arrow system, a researcher may study variables that influence students' achievement with data that includes the constructs of student engagement, teacher competence, and home support. The researcher uses regression analysis and computes a coefficient for the relations of each of these variables with achievement to estimate their distinct contributions to predicting achievement (Hilpert and Marchand 2018; Symonds et al. 2024). In a regression analysis, the assumption is that each of the three variables relates to achievement in a way that, at least in part, is independent from the other variables. By the statistical assumptions governing the regression analysis, the variables are required to have no significant relationship with each other (often referred to as non-collinearity). Clearly, such assumptions constitute an oversimplification of the interrelations of these variables: they influence each other and interact with each other in non-trivial ways. A supportive home would boost engagement, as would a competent teacher, and, in turn, engagement could facilitate familial support of the student (Saqr et al. 2023, 2024d). As such, assuming that variables are independent and unrelated does not reflect real-life social-psychological processes that underlie learning and achievement. Conceptualizing learning and achievement as a complex dynamic system, in which the three variables interdepend and mutually impact each other, would better correspond to reality (Saqr et al. 2024d; López-Pernas and Saqr 2024). A complex dynamic systems approach would allow conceptualizing, investigating, and describing the interactions between these variables, and the mutual influence they exert on each other in a way that reflects the multifaceted nature of learning and achievement. That, however, requires appropriate methods that are able to capture the dynamics, processes, interactions, and intricacies among these variables, and the way they give rise to the global behavior of the system. Such research would lead to a theory of the emergence patterns of the learning and achievement system, how the system's components reflect a process of self-organization, and how the system's global behavior constrains the state of the system's components and their interactions (Binkley et al. 2012; Garner and Kaplan 2019; Kaplan and Garner 2020; Scheffer 2020; Saqr et al. 2024d).

Thus, in contrast to box-and-arrow or component-dominant systems, complex dynamic systems are *interaction-dominant* - the interactions and the interdependencies between the components define system behavior. Hence, system behavior cannot be reduced to any of the components or their distinct interactions. As the system's components interact, new behavior, and behavior patterns emerge. Notably, the relationship between the system's behavior and the states of its components is not proportional, which manifests in nonlinear system behavior. Rather than an additive process, a complex system's components and their interactions give rise to behavior that is greater than the sum of its parts (e.g., planning + metacognition ≠ self-regulation). Complex systems are also not strictly predictable (Wallot and Kelty-Stephen 2018; Gernigon et al. 2024).

Adopting a CDS lens in education presents an opportunity to advance our existing theories and provide a deeper understanding of change, stability, and resilience in educational contexts. An opportunity to understand the dynamics and temporal aspects of how a learning process progresses, regresses or interacts with the context and environment. These dynamical aspects, while prevalent in our theories and practice, have remained hardly straightforward to capture with existing methods or interpret with existing theoretical frameworks. Recently, serious work has been done to renew theories within CDS see for instance (Symonds et al. 2024).

Using a CDS lens has resulted in significant breakthroughs in other fields e.g., Nobel prize in physics (Bianconi et al. 2023) and holds the promise for novel and meaningful insights to unravel the structure and temporal dynamics of learning. These opportunities not only promise a richer understanding but also offer practical implications for designing interventions that are adaptive, resilient, and responsive to the nuances of educational systems. To further explain the complex systems and their dynamics we will discuss the main characteristics and concepts of complex systems in the remainder of this section.

## Nonlinearity

Nonlinearity is a defining feature of complex systems, describing relationships between system components as not proportional or additive and highly sensitive to initial conditions (Favela and Amon 2023). Within a nonlinear system, the output (or behavior) of the system is not directly proportional to the input, where small changes can lead to disproportionately large outcomes, and vice versa, also described as a "butterfly effect" (Lorenz 2001). This concept directly contrasts with linear systems in which the system's output is easily predicted by the inputs, such as represented by the equation $y = mx + b$ and similar variations. In complex systems, nonlinear relationships can manifest in ways that defy conventional intuition ((e.g., Clark and Luis 2020)) where a small environmental disturbance, such as the interruption of a classmate, can trigger a cascade of effects that can dramatically and irrevocably alter learning processes, (e.g., Clark and Luis 2020) disruption of cognitive engagement. A typical example of non-linearity in educational psychology is the Yerkes-Dodson law, which describes the relationship between arousal and performance (Yerkes and Dodson 1908). The Yerkes-Dodson Law states that performance improves as physiological or mental arousal increases, but only to an optimal level. Beyond this point, excessive arousal leads to a decline in performance (see Figure 2).

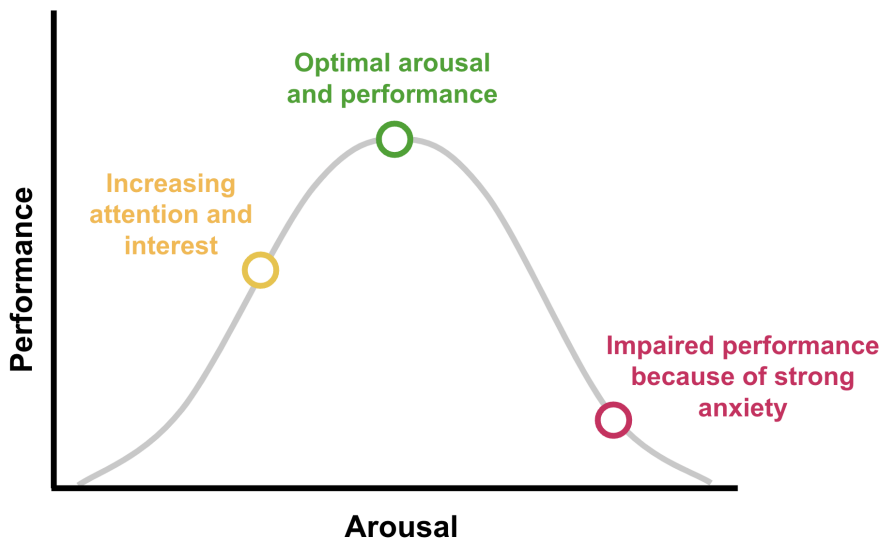

**Figure 2.** The Yerkes-Dodson Law (Yerkes and Dodson 1908) illustrates the relationship between arousal and performance. Performance increases with arousal up to an optimal point, beyond which excessive arousal impairs performance. The graph highlights three key zones: increasing performance due to heightened focus and motivation at low-to-moderate arousal (left), optimal performance at moderate arousal levels (peak), and declining performance caused by excessive stress and cognitive overload at high arousal levels (right).

Similarly within learning processes, initial conditions (e.g., the age of a pedagogical agent) can significantly and disproportionately affect students' learning outcomes. Nonlinearity in complex systems highlights that systems are not merely the sum of their parts; rather, the interactions between components of a system and the elicitation of these interactions by initial condition can create emergent properties, leading to unforeseeable outcomes by considering and examining individual components alone.

## Emergence and Self-organization

In complex systems, *emergence* refers to how new properties, patterns, and behaviors arise from the interactions between a system's individual components, although these behaviors are not present within the components themselves (Kalantari et al. 2020; Artime and De Domenico 2022). In other words, the behavior of the whole system cannot be attributed to a single component and the contribution of any component cannot be isolated, nor can it be added to any other putative distinct contributions. It is the interaction that elicits these behaviors. For example, in psychological phenomena such as self-regulated learning, the process of self-regulation cannot occur by just enacting individual components, or strategies. Instead, it is the interaction between these strategies, such as engaging in information processing while evaluating content as (ir)relevant, from which this process emerges (Dever et al. 2022). The emergence concept illustrates how the whole system becomes more than just the sum of its parts, exhibiting capabilities that transcend the faculties of individual components.

*Self-organization*, closely related to the concept of emergence, is the process by which order spontaneously arises from disorder without the oversight of a central controller (Camazine et al. 2003; Dever et al. 2023). Self-organization can be explained by the presence of simple local rules that lead to global-scale patterns of behaviors. Continuing our example in applying these concepts to self-regulated learning from above, students' self-regulation can follow a set of simple rules resulting in self-organization. For example, a simple rule could be that after defining a task, a student should engage in planning procedures. This rule can trigger a set of several cognitive and metacognitive processes (e.g., prior knowledge activation, planning, goal-setting) that occurs without the dictation of a central controller, representing self-organization in learning processes and other educational phenomena. Within psychological phenomena, this can be seen in group dynamics in which social norms, cultures, or identities are formed within groups and result in synchronized behavior, such as that seen within amusement parks or vehicular traffic. The self-organization of these systems is driven by feedback mechanisms that constrain and reinforce patterns from which behavior can demonstrate stability and adaptability (see Figure 3).

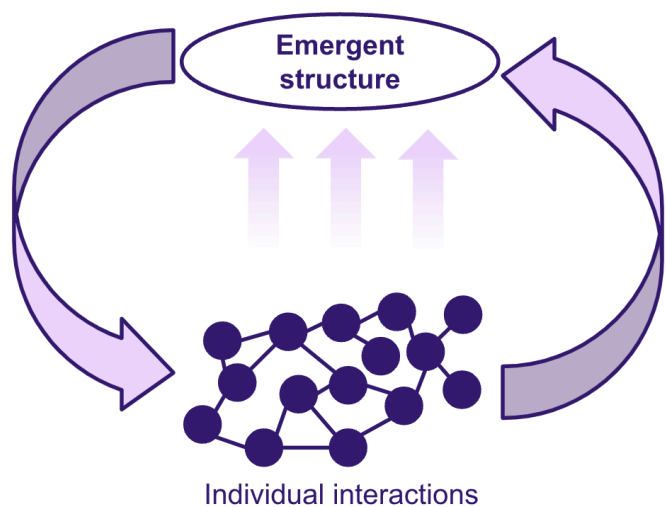

**Figure 3.** Individual interactions (bottom) collectively give rise to emergent properties and patterns (middle), which form higher-order structures through self-organization (top). These emergent structures, in turn, influence and shape the underlying interactions, creating a continuous feedback loop that drives the dynamic behavior of the system.

## Feedback Loops

A fundamental concept within complex systems, *feedback loops* occur when the output of a system feeds back into itself to serve as inputs. These feedback loops serve as the way in which a system amplifies its behavior (positive feedback loop) or stabilizes changes within the system (negative feedback loop), shaping the system's adaptability to environmental changes (Wiener 1948). Positive feedback loops accelerate change wherein the system moves away from its initial state and experiences exponential growth. We can think of positive feedback loops in terms of the engagement of cognitive and metacognitive strategies during learning. As students become increasingly engaged with learning materials, develop conceptual understanding, and mature in their abilities to adequately deploy learning processes, the student's cognition and behavior can move away from its initial state, destabilizing to further mature and develop in their application of learning processes.

Negative feedback loops moderate the change within the system to stabilize behaviors, leading to systems that demonstrate equilibrium. For example, stabilization of cognitive processes, such as in the repetition of learning strategies over a short period of time, can result in the equilibrium of a system in which the learning establishes a new "initial" state. In complex systems, the presence of both positive and negative feedback loops is essential for the balance and functionality of a system wherein excessive change can lead to unsustainable growth and subsequent collapse but a substantive lack of change indicates stagnation.

With psychological phenomena, such as emotion regulation, the balance of activating and deactivating states are essential to the stability and health of a system. In this context, an example of a positive feedback loop is the interaction between anxiety and procrastination (Fig. 4 - left): anxiety about a task leads to procrastination, which in turn increases anxiety as deadlines approach, creating a self-reinforcing cycle that can spiral into overwhelming anxiety (instability). On the other hand, a negative feedback loop can involve anxiety and self-regulated time-management strategies (Fig. 4 - right), where anxiety triggers the use of time-management techniques, such as breaking tasks into smaller steps or scheduling work sessions. These strategies reduce anxiety by promoting a sense of control and progress, gradually stabilizing emotional fluctuations and converging toward a balanced state of productivity.

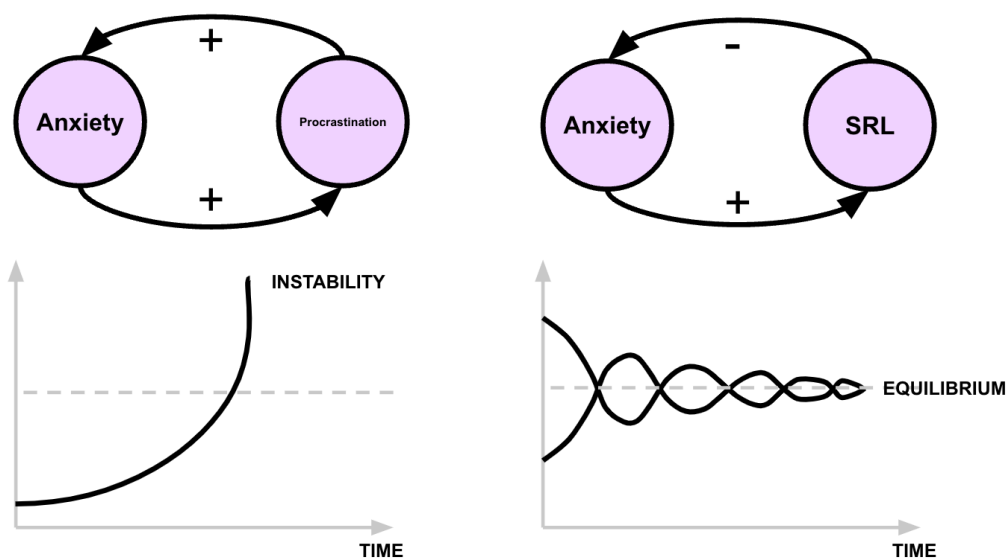

**Figure 4.** An example of a positive feedback loop (left) shows how anxiety and procrastination reinforce each other, amplifying anxiety and potentially leading to overwhelming stress (instability). In contrast, the negative feedback loop (right) demonstrates how anxiety triggers self-regulated time-management strategies, which reduce anxiety and stabilize emotional fluctuations over time, converging toward a balanced and productive state (equilibrium). Adapted from Rosnay (1979).

## Adaptation and Evolution

Feedback loops serve as the mechanisms by which a complex system can *adapt* and *evolve*. Through these mechanisms, a system can respond to both internal and external conditions wherein the components of a system can make decisions based on local information, self-organize into new configurations that respond to added conditions, and maintain functionality within dynamic environments (Ao et al. 2007; Wang et al. 2024). However, there is a balance that must be struck in the adaptability of a system in which a system that is too rigid or too flexible can result in counterintuitive behaviors that result in the instability of a system. Rigidity, broadly, refers to the inability for system behaviors to adapt to the external pressures and influences of the surrounding environment. In being too flexible, however, system behaviors do not adhere to the rules set to determine how the individual components react to environmental changes and in response to other system changes. For example, if a learners' self-regulation system is too rigid, this may be characterized by the increased repetition of minimal learning strategies such that a learner may only apply summarizing and note-taking to the learning materials. This presents a potential issue in that if the learning material no longer supports these two strategies (such as seen within non-traditional lectures and materials), the learner may not be able to apply more novel learning strategies. Conversely, if the learners' behaviors are too flexible and do not adhere to the internal rules of the system, the learner does not demonstrate structure which is imperative for learning.

The adaptivity of a system can result in the evolution of the same system where, because of the changes the system was required to undergo as a reaction to internal or environmental changes, the emergent properties and patterns of system dynamics no longer replicate those present within the original system (Corning 1995). This is modeled in systems that display increasingly more sophisticated and complex behaviors as time progresses. For example, evolution within educational psychological systems is determined by the emergence of new behaviors or abilities within students. In reference to psychological phenomena, educational psychologists have coined the term *learning* wherein the understanding of an individual changes as a result of new information present within the environment.

## Hierarchies, Scales, and Network Structures

Complex systems are characterized by their organization and interactions across organizational levels. This organization allows us to understand system structures and the influence each structure and connection has on each other for informing adaptive and evolutionary behaviors. *Hierarchical,* or multi-level structures, refer to a system's organization of its components in layers where each level consists of components that can be considered sub-components of a higher-level system (Siegenfeld and Bar-Yam 2019). Each of these levels can function on different spatial and temporal *scales* wherein the behavior of an individual component can affect a larger system. When examining complex systems, it is essential for scales to be considered as observing only one level may not allow for the patterns of the full system to be observed. For example, self-regulated learning consists of several macro-processes which each contain several micro-processes. Examining only a singular macro-process (e.g., planning) – while providing valuable insights about planning, a specific aspect of the self-regulatory process– narrows the scope in which the self-regulatory behavior cannot be fully examined. *Networks* are the frameworks that connect each component of a system which shapes how information is conveyed across the different components, whether the components are different levels or scales (Wang et al. 2024). Within a networked system, components are connected through relationships that dictate how components interact with each other. These relationships determine the speed at which information spreads, the resilience of a system to failures, the uniformity of system components in relation to each other, and the elicitation of emergent behaviors (Walker 2020). Because of these hierarchies, scales, and networks, it is difficult to determine the underlying causes of behaviors and predict future system behaviors but is vital for the adaptability and evolution of the complex system (Figure 5).

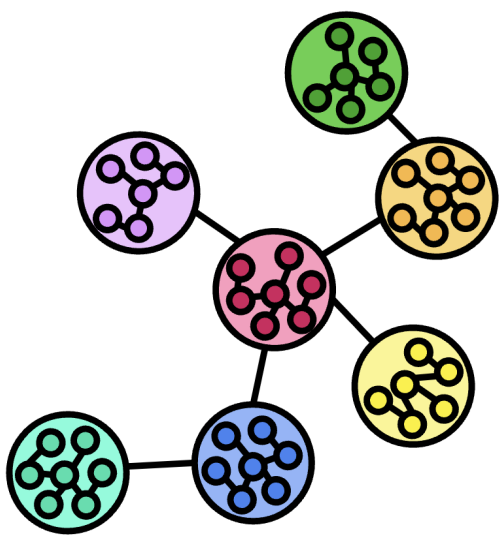

**Figure 5.** This network visualization illustrates the hierarchical or multi-level structure of a complex system, with each colored cluster representing a distinct domain of interconnected components. For example, one cluster could represent Self-Regulated Learning (SRL), including components such as goal-setting, self-monitoring, and reflection. Other clusters represent related domains like Engagement and Motivation, each with their own interconnected sub-components. The links between clusters represent the interactions between these domains, demonstrating how a change in one area, such as SRL, can cascade and influence others, like Motivation or Engagement. The thickness in the links between clusters may indicate the strength of the relationship between clusters where colors (e.g., green versus red) may indicate the positive or negative relationship between the clusters.

## Dynamics and Sensitivity to Initial Conditions

Complex systems exhibit dynamic behaviors that arise from the interactions between system components. System *dynamics* refer to the changes exhibited by the system over time due to the evolution of component interactions and the influence of external conditions from the environment. The dynamics of a system can demonstrate cycles of repeated behaviors, chaotic behaviors that appear random, or emergent in which unforeseen patterns suddenly or gradually arise from local interactions (Favela and Amon 2023). However, these dynamics are greatly sensitive to initial conditions in which small variations in the starting point of a process can have a multiplicative effect on system behaviors later on. This ties into the aforementioned "butterfly effect" concept mentioned earlier in this section. In all, the sensitivity to initial conditions makes it nearly impossible to employ predictive modeling when encountering complex systems where short-term behavior may be relatively predictable in contrast to long-term behaviors due to the amplification over a temporal scaling. *Attractor states* serve as the stable configuration of relationships and behaviors that a complex system tends to evolve toward (Han and Amon 2021). This can be thought of as the 'destination' of the system in which a mathematical model aims to represent all possible states that a complex system occupies, including the intermediary states, on its course towards a convergence point (see Figure 6). An example is found in the study by Gao et al. (2024) exploring students' motivation, where the authors found that several initial conditions interact closely with academic affordances, steering the system toward attractor states primarily influenced by career planning.

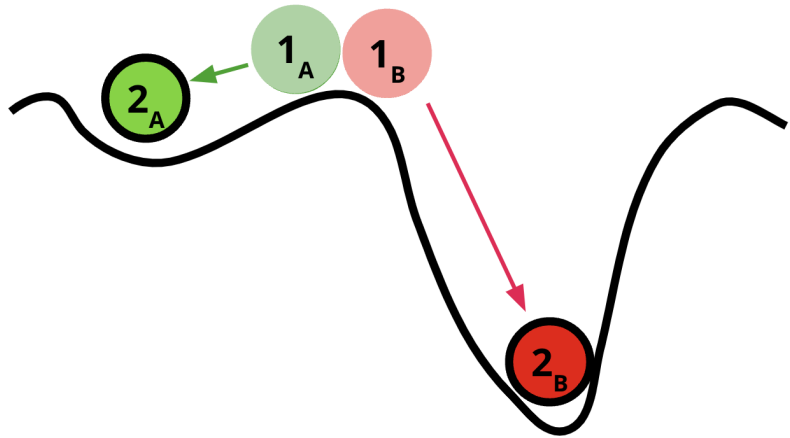

**Figure 6.** Complex systems tend to evolve toward **attractor states**—stable patterns or behaviors that the system naturally gravitates toward. These attractors can reflect positive stability, where the system maintains a desirable equilibrium ($2_A$), or, on the other hand, they can signify being "stuck", where the system is locked into a maladaptive or undesirable state that resists change ($2_B$). The system's initial conditions ($1_A$ vs. $1_B$) significantly influences its future behavior; tiny differences at the start can result in vastly different outcomes and lead to evolve towards a positive attractor state ($1_A$ to $2_A$) or to a negative one ($1_B$ to $2_B$).

# Analytical approaches

Scientific paradigms shape the nature of research questions and the methods used to answer them (Kuhn 1962). The complex systems approach uses certain methods of investigation and has developed its own toolbox of analysis and computer modeling.

# Network analysis: Capturing the interactions and dynamics

Network analysis is a quantitative method to identify the structure and interactions between components of a complex system (Saqr et al. 2022, 2024b). Networks are represented as nodes (i.e., vertices) and edges (i.e., links) where nodes typically represent the component of a complex system and the edges represent the relationships between these components. These nodes and edges serve as the fundamental building blocks of networks, identifying intersecting relationships between system components. For example, if we take self-regulated learning (see Figure 7) as a complex system, a specific strategy (such as goals setting) will be represented as a node where edges will be visible to show how strategies connect with each other, e.g., regulation, or help-seeking (Saqr and López-Pernas 2024). This analytical approach is extremely valuable for capturing and examining interdependent relationships wherein it is essential to dissect the degree to which nodes are related to each other.

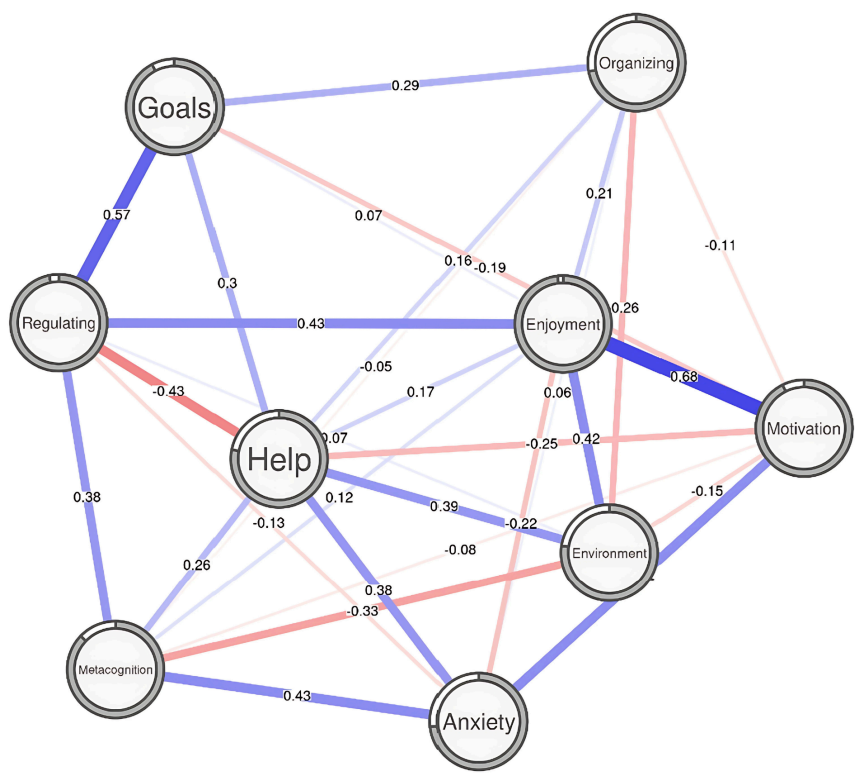

**Figure 7.** Partial correlation network between SRL components (Saqr and López-Pernas 2024). The thickness of each edge is proportional to the strength of partial correlation. Higher values indicate stronger partial correlation and therefore interdependence.

From this analytic method, researchers can capture the number of connections each node has which quantifies the degree of influence a particular node has on network dynamics (i.e., strength or expected influence), if a node acts as a critical connector for other parts of a system or other networks which reveals its role in transferring information within a system (i.e., betweenness centrality), the tendency of a system to form tight sub-groups which reveals system hierarchies and interconnectedness (i.e., clustering), and the distance between nodes which reveals the efficiency of information spread across a network (i.e., path length; Brass and Borgatti 2019). These network analytics can allow researchers to visualize system connectedness, patterns, and outliers, analyze how these relationships change over time, and identify the complexity of the system structure (e.g., nested, hierarchical, multilayered) (Malmberg et al. 2022; Saqr et al. 2024a).

There are several examples in the literature of the use of psychological networks to model psychological or education processes. SRL, as exemplified before, has been one of the phenomena most commonly modeled through complexity methods. For instance, López-Pernas et al. (2025) used partial correlation networks (Saqr et al. 2024a) to model the relationship between students' SRL components. The authors found that the SRL dimensions with the most influence in the SRL process differ from the beginning (when metacognition is more influential) to the end of the course (motivation), and therefore might be more amenable to intervention. Saqr & López-Pernas (2024) used different temporal and contemporaneous networks to study the differences between group-based (depicted in Fig. 6) and individual SRL dynamics. The authors concluded that the average SRL process is not representative of the individual SRL processes of each student, advocating for the study of idiographic analysis to gather personalized insights. Using a similar approach, Saqr (2024) studied the interplay of engagement dimensions using log data from a learning management system and found marked differences between group-level and within-person variance. These highlight that complex systems behave differently for different individuals, depending on the initial conditions, the interdependence between the system's components, and the influence of external variables.

Through a transition network analysis approach (Saqr et al. 2024c), López-Pernas and Saqr (2024) studied the longitudinal dynamics of online engagement using sequence analysis and transition networks (Fig. 8-left) and found that engagement patterns remain relatively stable over time, aligning with the typical behaviors observed in complex systems (i.e., attractor states, Fig. 8-right). Significant changes in these patterns are rare, and when they do occur, they are generally brief, with students often reverting to their original engagement levels. This suggests that a student's initial engagement state heavily influences their long-term trajectory, effectively determining their likely outcomes.

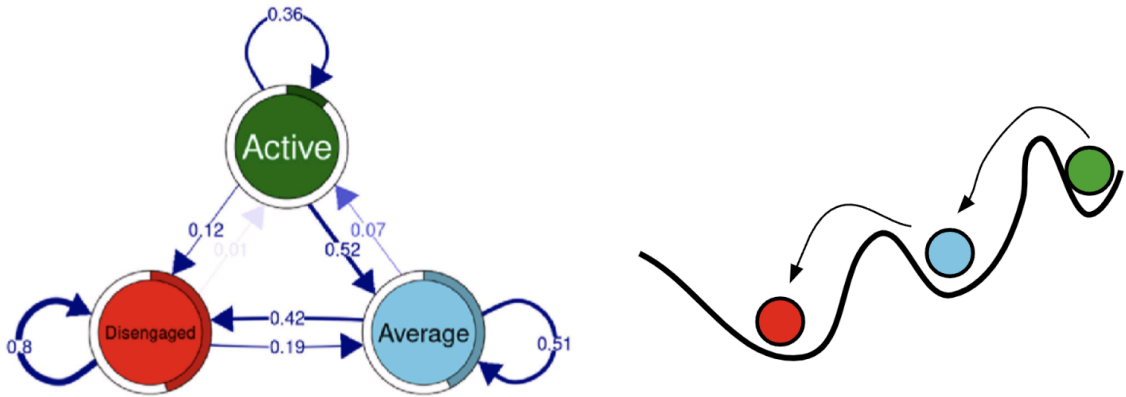

**Figure 8.** Transition network of students in the *mostly disengaged* trajectory identified by (2024). Students in this trajectory tend to reach a disengaged state and remain disengaged.

Lastly, it is worth mentioning that, although SRL and engagement can be studied as complex systems of their own, they can be also modeled as part of a macro complex system where they interact with one another and other components (e.g., motivation), as proposed by (Saqr et al. 2024a) (Fig. 9).

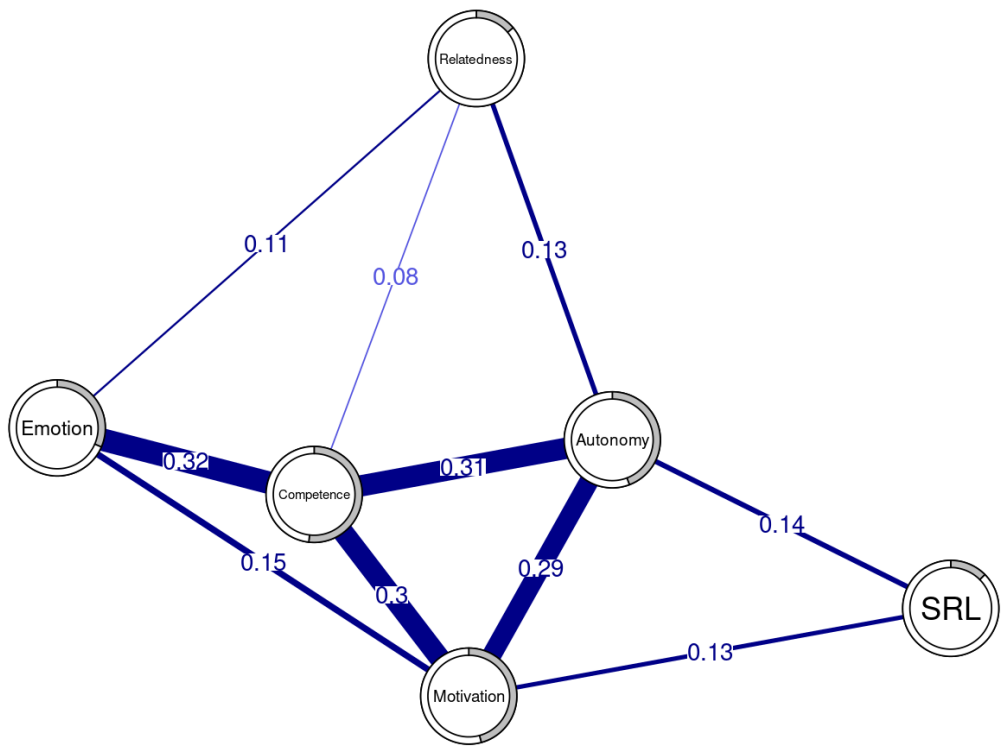

**Figure 9.** Psychological network illustrating the complex system formed by SRL and other constructs (Saqr et al. 2024a)

# Recurrence Quantification Analysis

Recurrence Quantification Analysis (RQA) is a statistical technique that reveals the nonlinearity and dynamic properties of time series data (Webber and Zbilut 2004; Webber et al. 2009). This method detects and quantifies how systems shift between repetitive and novel sequences of behaviors across a temporal space. Three types of RQA methods exist: auto-RQA (aRQA), cross-RQA (cRQA), and multidimensional-RQA (mdRQA). aRQA examines the dynamics within a singular time series, cRQA identifies shared dynamics of two time series, and mdRQA identifies the repeating patterns across more than two time series (Moulder et al. 2023). These analyses can be completed with either categorical or continuous time series data which are mapped against each other on a matrix where Time Series 1 is mapped on the x-axis and Time Series 2 is mapped on the y-axis. On the matrix, the intersection of time series values are highlighted black if the values are considered the same, pre-determined using radial metrics (this is explained in the tutorial down below).

Taking categorical aRQA as an example, Time Series 1 [A, B, X, Y, A, S, X, Y] will be the same dataset as Time Series 2 [A, B, X, Y, A, S, X, Y]. Every time on the matrix time points match in value, such as that with time points t1 and t5 as well as t3 and t7, the intersection on the matrix is shaded black (see Figure 10). The diagonal line represents the line of identity (LOI) where the time series will always be recurrent with itself at Lag 0. Lags are defined by the distance between time points. For example, Lag 1 represents time points that are removed by one time space (e.g., Time 1 and Time 2, Time 2 and Time 3); Lag 2 represents time points removed by two time spaces (e.g., Time 1 and Time 3). By incorporating lags into subsequent analyses, RQA considers the temporal relationships between events such that patterns in data can emerge across large time differences.

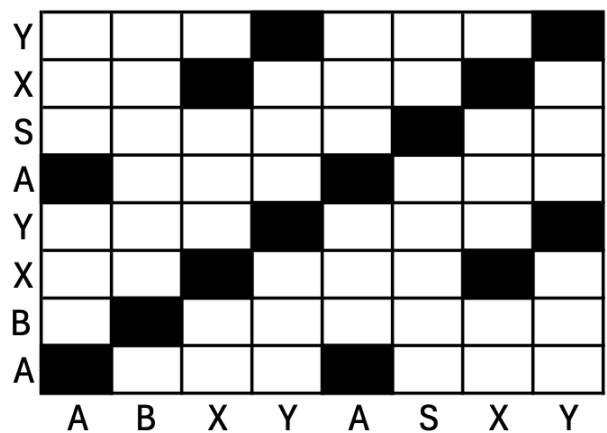

**Figure 10.** Example of aRQA recurrence matrix.

RQA outputs several metrics, including recurrence rate (proportion of repetition), percent determinism (proportion of patterns on diagonals), average diagonal line length (average length of diagonal line structures), laminarity (proportion of repetitions on vertical line structures), and trapping time (average length of vertical lines). This is not an exhaustive list, but it represents the vast amount of information that can be extracted through this analysis that can explain the degree to which a system displays dynamism, how events that occur within the environment impact the dynamics of the system, and the degree to which a system demonstrates adaptivity. For example, a study by Dever et al. (2022) applied this analytical methodology to self-regulated learning strategies during game-based learning. In identifying significant differences in the metrics output by aRQA analyses according to learning outcomes and agency experienced by students in the game, Dever et al. (2022) established how students demonstrated functionality in their self-regulation during learning and were able to identify instances in which students demonstrated dysfunctional self-regulation. This provides interesting insights into how students should be supported and scaffolded within their self-regulated learning behaviors. For example, can we identify when a student demonstrates dysfunctional behaviors and how can we design interventions which help the student to correct this dysfunction to demonstrate functional self-regulation?

## Simulation

Nowadays, computers make it possible to implement complex systems of equations of change. But for those who shy away from mathematical formalization, there are user-friendly computer programs that make it very easy to simulate the complexity of interactions at the origin of self-organization processes and attractor dynamics (Nowak 2004; Vallacher 2017; Gernigon et al. 2024). Among the most popular computer simulation tools are dynamic networks, agent-based models, and cellular automata. As explained earlier in this chapter, a dynamic network models a complex system as a network of connected nodes (the system's components). This type of computational tool has been used to model and simulate the dynamics of several phenomena of interest to the educational field, including intelligence development (van der Maas et al. 2006), language development (van Geert 1991), parent-child interactions (van Dijk et al. 2013), and goal-directed motivation (Westaby et al. 2014). One of the most user-friendly open-source software packages for creating dynamic network models is Insight Maker[1]. Agent-based models simulate the behavior of virtual agents that represent the components of a system. These agents interact with each other and with their virtual environment based on iteratively applied evolution rules. Agent-based models have thus made it possible to model the dynamics of phenomena important to the educational context, such as motivation (Jager et al. 2001; Merrick and Shafi 2013), language development (Steels and Belpaeme 2005), and parent-child interactions (Hesp et al. 2019). One of the most widely used open-source, user-friendly programming platforms for agent-based modeling is NetLogo[2]. Finally, cellular automata are simplified forms of agent-based models, where agents are the cells of a two-dimensional lattice or a line. The state of each cell at time t+1 depends, through the application of an evolution rule, on the state of the cells in its neighborhood at time t. The same rule is then applied to the new generation of cells at t+1 to shape generation t+2, and so forth. One of the most insightful applications of cellular automata to psychological development processes concerns the self-organizing dynamics of the self-concept into differentiated structures (Nowak 2004), that is, into distinct clusters that may virtually represent different dimensions of the self, such as the academic self, the athletic self, the relational self, etc. Among the open-source platforms for user-friendly implementation of cellular automata we might mention Golly[3].

---

[1] Insight Maker: https://www.insightmaker.com
[2] NetLogo: https://ccl.northwestern.edu/netlogo

Unlike the component-dominant mainstream approach, which aims to identify causal factors, the complex dynamic systems paradigm focuses on the identification of interaction-dominant processes. This is what the research methods we have just briefly reviewed contribute to, by aiming to experimentally test the volatility and stability of systems' behavior, to identify statistical signatures of complexity and non-linearity in the time series of these systems' states, to formally and/or computationally model the self-organization processes at work within them, and, for the sake of validating the models built, to retrieve the same statistical signatures in the time series generated by their computer simulation. Once validated, these dynamic models are invaluable tools for testing the effects of interventions, such as pedagogical ones, on the dynamics of important educational variables.

# Data collection

## Capturing a system's state through rich description: interviews and written reflections

Understanding complex phenomena, such as individuals' learning, motivation, identities, and actions, requires methods that capitalize on the richness and depth of personal experiences. Narrative interviews and written reflections are markedly effective for capturing the state of a complex system at specific points in time. These methods provide insight into how individuals interpret and enact their roles, goals, beliefs, perceptions, and learning strategies within their unique lived contexts (Kaplan and Garner).

Narrative interviews serve as an effective tool for exploring individuals' experiences and meaning-making processes. Interviewers prompt participants to share personal stories, revealing the interplay between system elements such as beliefs, goals, self-perceptions, emotions, and actions. For example, research on teachers' professional development has used interviews conducted before, during, and after the program to trace changes in participants' motivations and instructional practices (Garner and Kaplan 2019). These interviews allowed participants to reflect on their teaching journey, the influence of prior experiences, and how professional

---

[3] Golly: https://golly.sourceforge.io

development shaped their beliefs about student learning and instructional strategies.

Written reflections provide another method for capturing detailed and introspective data. These reflections can be gathered over time, such as through daily or weekly diary entries, offering snapshots of participants' evolving thoughts and decisions. For instance, teachers reflecting on their experiences with inquiry-based learning might describe how their beliefs about teaching shifted during the professional development program. Such reflections illuminate the dynamic relationships among the person's system components, revealing how changes in context or new experiences challenge or reinforce existing beliefs, goals, and strategies.

The analysis of these rich data sources involves identifying key elements of the individual's complex system, such as their beliefs, values, goals, emotions, and actions, as well as the relationships among them (Malmberg et al. 2022; Törmänen et al. 2023). Researchers synthesize these findings to create a holistic understanding of the individual's system at specific points in time, highlighting patterns and tensions within and across system elements. Additionally, comparing data across participants allows for researchers to identify shared patterns and generate theoretical principles that explain how identities and actions develop in different contexts.

These methods emphasize the importance of capturing the nuanced and context-dependent nature of complex systems. Narrative interviews and written reflections provide a deep and detailed understanding of individuals' experiences, enabling researchers to document the states of their systems at key moments and track their evolution over time.

## Capturing real-time emergence through observations and stimulated recall

Studying real-time emergence in complex systems requires methods that capture dynamic processes as they unfold. Observations and stimulated recall interviews are particularly effective for examining how engagement emerges within specific contexts, providing detailed accounts of participants' experiences, interactions, decisions, and actions (Kaplan and Garner).

Detailed observations allow researchers to document participants' behaviors and interactions in naturalistic or designed settings. For example, in a study of visitors to a museum exhibition designed to promote inventiveness, the researchers used observations to capture participants' engagement with the exhibition's physical and conceptual features. Observational data revealed how visitors interacted with the content, expressed their perceptions, and responded to the contextual elements of the exhibition. This real-time data provided insight into how participants' identities and goals were activated and shaped by the environment (Kaplan et al. 2023).

Stimulated recall interviews build on observational data by eliciting participants' reflections on their thoughts, emotions, and decisions during specific moments of activity. Conducted shortly after the observed session, these interviews prompt participants to describe their experiences in detail, often using notes or video footage as memory aids. For example, in the museum study described above, following an experience with an artifact, participants were asked to recount their thought processes during the visit, providing insights into how they connected the exhibition content to their roles and goals, such as seeing themselves as creative or inventive in other areas of life.

Combining observations with stimulated recall interviews allows researchers to construct a comprehensive picture of how engagement emerges in real-time through participants' experiences in an authentic educational context. Observations provide a foundation for understanding what participants do and say within the context, while stimulated recall interviews add depth by uncovering the subjective meanings behind these actions. Analyzing these data sources together allows researchers to identify patterns of interaction between contextual features and individual dispositions, highlighting the dynamic and context-dependent nature of engagement emergence.

These methods are particularly valuable for capturing the temporal and situational aspects of complex systems. Observations provide a detailed record of real-time activity, while stimulated recall interviews offer access to participants' reflective processes. Together, these approaches enable researchers to study the dynamic interplay between individuals and their environments, shedding light on how identities and actions develop and transform within specific contexts.

# Final remarks

The complex dynamic systems approach lends itself to empirical observation, but in a different way to traditional scientific approaches. While the latter aim to discover —by isolating— causal relationships between independent and dependent variables, the complex dynamic systems approach investigates the way in which the system's global behavioral patterns, or order parameters, change or stabilize under the effect of a parameter external to the system, or control parameter, which constrains the system's internal interaction dynamics (Haken et al. 1985). For example, with regard to motor learning, Zanone and Kelso (1992) showed how an oscillatory synchronization motor skill to be acquired can emerge and then stabilize by training it to resist increasing perturbations. Here, perturbations play the role of a control parameter that is experimentally manipulated —by gradually increasing the required frequency of limbs' oscillation— in order to push the system of the emergent synchronization into disorganization. The resulting learning is thus defined in dynamic terms, that is, as the stabilization of the system's new behavior in the form of its resistance to perturbation.

The states of complex dynamic systems evolve over time while exhibiting typical statistical properties. Recurrence Quantification Analysis (RQA) can detect regularities and forms of stability (attractors) in the dynamics of the states of a complex system. The temporal distribution of these states also reveals a long-term memory property that translates into a particular autocorrelation structure that can be found whatever the time scale on which these states are observed. This statistical structure, which reveals dynamics based on the traces of its own past history, is a power-law temporal distribution known as $1/f$ noise or pink noise. Its detection is enabled by time-series analysis of the states of the system under study, particularly through Detrended Fluctuation Analysis (DFA; Peng et al. 1993) and the various adaptations of this technique that have been created to overcome its limitations (see, for example, the chapter by Altamore et al., this book). In the educational field, distributions in the form of $1/f$ noise have been identified in language learning (Lowie et al. 2014) and motor learning (Nourrit-Lucas et al. 2015), attesting that learning is indeed a dynamic phenomenon in the sense of the theory of complex dynamic systems. Interestingly, $1/f$ distributions are ubiquitous in nature (Bak 1999), and their replacement by other types of distribution often reflects anomalies (Van Orden 2007). Consequently, from an applied perspective, such replacements can be used as warning signals of a risk of deteriorating behavior, for example, a risk of discouragement towards academic studies or even academic failure. Detecting such a signal can then be very useful in deciding whether to implement individualized support procedures for students in difficulty, before it is too late.

Variations in the states of a complex dynamic system are the result of iterative processes that can be modeled mathematically. The momentary state of a system depends to a large extent on its previous state, which in turn depended on an earlier state, and so forth. When time is expressed in continuous form, this dependence of a system on its own history can be accounted for by differential equations such as $dx/dt = f(x)$, where d represents the variation in the system's state $x$ or the variation of time $t$. When time is discretized, history dependence is expressed by different equations such as $x_{t+1} = f(x_t)$. However, state variations of complex dynamic systems also result from iterative interactions (often reciprocal in the form of couplings) between their components. The above-mentioned equations of change are then enriched by the influences of changes in the variables interacting within the system, taking the form of systems of differential or difference equations. All these equations are capable of producing monostable (one attractor), multistable (several attractors that can be visited more or less cyclically), or even chaotic behaviors of the system. In the educational field, such equations of change have been used to model the dynamics of cognitive development (van Geert 1998), language development (van Geert 1991), social development (Steenbeek and van Geert 2008), and teaching-learning interactions (Merlone et al. 2019).

# References

Ao P, Kwon C, Qian H (2007) On the existence of potential landscape in the evolution of complex systems. Complexity 12:19–27. https://doi.org/10.1002/cplx.20171

Artime O, De Domenico M (2022) From the origin of life to pandemics: emergent phenomena in complex systems. Philos Trans A Math Phys Eng Sci 380:20200410. https://doi.org/10.1098/rsta.2020.0410

Bak P (1999) How Nature Works: the science of self-organized criticality, 1st edn. Springer, New York, NY

Bianconi G, Arenas A, Biamonte J, et al (2023) Complex systems in the spotlight: next steps after the 2021 Nobel Prize in Physics. J Phys Complex 4:010201. https://doi.org/10.1088/2632-072X/ac7f75

Binkley M, Erstad O, Herman J, et al (2012) Assessment and Teaching of 21st Century Skills. Assessment and Teaching of 21st Century Skills 1–15. https://doi.org/10.1007/978-94-007-2324-5

Brass DJ, Borgatti SP (2019) Social networks at work. Routledge

Bringmann L, Helmich M, Eronen M, Voelkle M (2023) Complex systems approaches to psychopathology. In: Krueger RF, Blaney PH (eds) Oxford Textbook of Psychopathology. Oxford University PressNew York, pp 103–122

Camazine S, Deneubourg J-L, Franks NR, et al (2003) Self-organization in biological systems. Princeton University Press, Princeton, NJ

Clark TJ, Luis AD (2020) Nonlinear population dynamics are ubiquitous in animals. Nat Ecol Evol 4:75–81. https://doi.org/10.1038/s41559-019-1052-6

Corning PA (1995) Synergy and self‑organization in the evolution of complex systems. Syst Res 12:89–121. https://doi.org/10.1002/sres.3850120204

Dever DA, Amon MJ, Vrzáková H, et al (2022) Capturing Sequences of Learners' Self-Regulatory Interactions With Instructional Material During Game-Based Learning Using Auto-Recurrence Quantification Analysis. Front Psychol 13:813677. https://doi.org/10.3389/fpsyg.2022.813677

Dever DA, Sonnenfeld NA, Wiedbusch MD, et al (2023) A complex systems approach to analyzing pedagogical agents' scaffolding of self-regulated learning within an intelligent tutoring system. Metacognition and Learning 18:659–691. https://doi.org/10.1007/s11409-023-09346-x

Favela LH, Amon MJ (2023) Enhancing Bayesian approaches in the cognitive and neural sciences via complex dynamical systems theory. Dynamics 3:115–136. https://doi.org/10.3390/dynamics3010008

Gao Y, Wang X, Fan P (2024) Exploring male English major's motivation trajectory through complex dynamic systems theory. Curr Psychol 43:9089–9100. https://doi.org/10.1007/s12144-023-05062-6

Garner JK, Kaplan A (2019) A complex dynamic systems perspective on teacher learning and identity formation: an instrumental case. Teach Teach 25:7–33. https://doi.org/10.1080/13540602.2018.1533811

Gernigon C, Den Hartigh RJR, Vallacher RR, van Geert PLC (2024) How the Complexity of Psychological Processes Reframes the Issue of Reproducibility in Psychological Science. Perspect Psychol Sci 18:952–977. https://doi.org/10.1177/17456916231187324

Gouvea J (2023) Processing misconceptions: dynamic systems perspectives on thinking and learning. Front Educ 8.: https://doi.org/10.3389/feduc.2023.1215361

Haken H, Kelso JA, Bunz H (1985) A theoretical model of phase transitions in human hand movements. Biol Cybern 51:347–356. https://doi.org/10.1007/bf00336922

Han J, Amon MJ (2021) A nonlinear dynamical systems approach to emotional attractor states during media viewing. CogSci 43:

Hesp C, Steenbeek HW, van Geert PLC (2019) Socio-Emotional CONcern DynamicS in a model of real-time dyadic interaction: Parent-child play in autism. Front Psychol 10:1635. https://doi.org/10.3389/fpsyg.2019.01635

Hilpert JC, Marchand GC (2018) Complex Systems Research in Educational Psychology: Aligning Theory and Method. Educ Psychol 53:185–202. https://doi.org/10.1080/00461520.2018.1469411

Jager W, Popping R, van de Sande H (2001) Clustering and Fighting in Two- party Crowds: Simulating the Approach-avoidance Conflict

Kalantari S, Nazemi E, Masoumi B (2020) Emergence phenomena in self-organizing systems: a systematic literature review of concepts, researches, and future prospects. J Organ Comput 30:224–265. https://doi.org/10.1080/10919392.2020.1748977

Kaplan A, Garner JK (2020) Steps for applying the complex dynamical systems approach in educational research: A guide for the perplexed scholar. J Exp Educ 88:486–502. https://doi.org/10.1080/00220973.2020.1745738

Kaplan A, Garner JK

Kaplan A, Garner J, Neuber A (2023) Learning as identity change, self-regulated learning as identity exploration: A complex dynamic systems perspective on the goals of education in the 21st century. In: Proceedings of the *1st World Giftedness Center International Conference*

Koopmans M (2020) Education is a Complex Dynamical System: Challenges for Research. J Exp Educ 88:358–374. https://doi.org/10.1080/00220973.2019.1566199

Kuhn T (1962) The structure of scientific revolutions. University of Chicago Press, Chicago, IL

López Pernas S, Conde-González M, Raspopović Milić M, Saqr M (2025) Frequencies and averages miss the point of SRL evolution: A complex dynamic systems approach. In: Proceedings TEEM 2024: Twelfth International Conference on Technological Ecosystems for Enhancing Multiculturality. TEEM 2024. Lecture Notes in Education Technology. Springer, Berlin, Germany, p in–press

López-Pernas S, Saqr M (2024) How the dynamics of engagement explain the momentum of achievement and the inertia of disengagement: A complex systems theory approach. Comput Human Behav 153:108126. https://doi.org/10.1016/j.chb.2023.108126

Lorenz E (2001) The butterfly effect. In: The Chaos Avant-Garde. WORLD SCIENTIFIC, pp 91–94

Lowie W, Plat R, de Bot K (2014) Pink noise in language production: A nonlinear approach to the multilingual lexicon. Ecol Psychol 26:216–228. https://doi.org/10.1080/10407413.2014.929479

Malmberg J, Saqr M, Järvenoja H, Järvelä S (2022) How the monitoring events of individual students are associated with phases of regulation: A network analysis approach. Journal of learning analytics 9:77–92

Merlone U, Panchuk A, van Geert P (2019) Modeling learning and teaching interaction by a map with vanishing denominators: Fixed points stability and bifurcations. Chaos Solitons Fractals 126:253–265. https://doi.org/10.1016/j.chaos.2019.06.008

Merrick KE, Shafi K (2013) A game theoretic framework for incentive-based models of intrinsic motivation in artificial systems. Front Psychol 4:791. https://doi.org/10.3389/fpsyg.2013.00791

Moulder R, Booth B, Abitino A, D'Mello S (2023) Recurrence quantification analysis of eye gaze dynamics during team collaboration. In: LAK23: 13th International Learning Analytics and Knowledge Conference. ACM, New York, NY, USA

Nourrit-Lucas D, Tossa AO, Zélic G, Delignières D (2015) Learning, motor skill, and long-range correlations. J Mot Behav 47:182–189. https://doi.org/10.1080/00222895.2014.967655

Nowak A (2004) Dynamical minimalism: why less is more in psychology. Pers Soc Psychol Rev 8:183–192. https://doi.org/10.1207/s15327957pspr0802_12

Olthof M, Hasselman F, Oude Maatman F, et al (2023) Complexity theory of psychopathology. J Psychopathol Clin Sci 132:314–323. https://doi.org/10.1037/abn0000740

Panadero E (2017) A Review of Self-regulated Learning: Six Models and Four Directions for Research. Front Psychol 8:422. https://doi.org/10.3389/fpsyg.2017.00422

Peng C-K, Mietus J, Hausdorff JM, et al (1993) Long-range anticorrelations and non-Gaussian behavior of the heartbeat. Phys Rev Lett 70:1343–1346. https://doi.org/10.1103/PhysRevLett.70.1343

Rosnay J (1979) The macroscope: A new world scientific system. NY Harper Row, New York

Saqr M (2024) Group-level analysis of engagement poorly reflects individual students' processes: Why we need idiographic learning analytics. Comput Human Behav 150:107991. https://doi.org/10.1016/j.chb.2023.107991

Saqr M, Beck E, López-Pernas S (2024a) Psychological Networks: A Modern Approach to Analysis of Learning and Complex Learning Processes. In: Saqr M, López-Pernas S (eds) Learning Analytics Methods and Tutorials: A Practical Guide Using R. Springer Nature Switzerland, Cham, pp 639–671

Saqr M, López-Pernas S (2024) Mapping the self in self-regulation using complex dynamic systems approach. Br J Educ Technol. https://doi.org/10.1111/bjet.13452

Saqr M, López-Pernas S, Conde-González MÁ, Hernández-García Á (2024b) Social network analysis: A primer, a guide and a tutorial in R. In: Learning Analytics Methods and Tutorials. Springer Nature Switzerland, Cham, pp 491–518

Saqr M, López-Pernas S, Helske S, Hrastinski S (2023) The longitudinal association between engagement and achievement varies by time, students' subgroups, and achievement state: A full program study. Comput Educ 104787

Saqr M, López-Pernas S, Törmänen T, et al (2024c) Transition network analysis: A novel framework for modeling, visualizing, and identifying the temporal patterns of learners and learning processes. arXiv [cs.SI]

Saqr M, Poquet O, Lopez-Pernas S (2022) Networks in education: A travelogue through five decades. IEEE Access 10:32361–32380. https://doi.org/10.1109/access.2022.3159674

Saqr M, Schreuder MJ, López-Pernas S (2024d) Why Educational Research Needs a Complex System Revolution that Embraces Individual Differences, Heterogeneity, and Uncertainty. In: Saqr M, López-Pernas S (eds) Learning Analytics Methods and Tutorials: A Practical Guide Using R.

Springer Nature Switzerland, Cham, pp 723–734

Sayama H (2015) Introduction to the Modeling and Analysis of Complex Systems. Open SUNY Textbooks

Scheffer M (2020) Critical Transitions in Nature and Society. Princeton University Press

Siegenfeld AF, Bar-Yam Y (2019) An introduction to complex systems science and its applications. arXiv [physics.soc-ph]

Steels L, Belpaeme T (2005) Coordinating perceptually grounded categories through language: a case study for colour. Behav Brain Sci 28:469–89; discussion 489–529. https://doi.org/10.1017/S0140525X05000087

Steenbeek H, van Geert P (2008) An empirical validation of a dynamic systems model of interaction: do children of different sociometric statuses differ in their dyadic play? Dev Sci 11:253–281. https://doi.org/10.1111/j.1467-7687.2007.00655.x

Symonds JE, Kaplan A, Upadyaya K, et al (2024) Momentary student engagement as a dynamic developmental system. J Theor Philos Psychol. https://doi.org/10.1037/teo0000288

Thurner S, Hanel R, Klimek P (2018) Introduction to the theory of complex systems. Oxford University Press, London, England

Törmänen T, Järvenoja H, Saqr M, et al (2023) Affective states and regulation of learning during socio-emotional interactions in secondary school collaborative groups. Br J Educ Psychol 93 Suppl 1:48–70. https://doi.org/10.1111/bjep.12525

Vallacher RR (2017) Computational Social Psychology. Routledge, New York : Routledge, 2017. | Series: Frontiers of social psychology

van der Maas HLJ, Dolan CV, Grasman RPPP, et al (2006) A dynamical model of general intelligence: the positive manifold of intelligence by mutualism. Psychol Rev 113:842–861. https://doi.org/10.1037/0033-295X.113.4.842

van Dijk M, van Geert P, Korecky-Kröll K, et al (2013) Dynamic adaptation in child–adult language interaction. Lang Learn 63:243–270. https://doi.org/10.1111/lang.12002

van Geert P (1991) A dynamic systems model of cognitive and language growth. Psychological Review 98:3–53. https://doi.org/10.1037/0033-295X.98.1.3

van Geert P (1998) A dynamic systems model of basic developmental mechanisms: Piaget, Vygotsky, and beyond. Psychol Rev 105:634–677. https://doi.org/10.1037/0033-295x.105.4.634-677

Van Orden G (2007) The fractal picture of health and wellbeing. PsycEXTRA Dataset

Walker BH (2020) Resilience: what it is and is not. Ecol Soc 25.: https://doi.org/10.5751/es-11647-250211

Wallot S, Kelty-Stephen DG (2018) Interaction-dominant causation in mind and brain, and its implication for questions of generalization and replication. Minds Mach (Dordr) 28:353–374. https://doi.org/10.1007/s11023-017-9455-0

Wang J, Zhang Y-J, Xu C, et al (2024) Reconstructing the evolution history of networked complex systems. Nat Commun 15:2849. https://doi.org/10.1038/s41467-024-47248-x

Webber CL Jr, Marwan N, Facchini A, Giuliani A (2009) Simpler methods do it better: Success of Recurrence Quantification Analysis as a general purpose data analysis tool. Phys Lett A 373:3753–3756. https://doi.org/10.1016/j.physleta.2009.08.052

Webber C, Zbilut J (2004) Recurrence quantification analysis of nonlinear dynamical systems. In: M. A. Riley GVO (ed) *Tutorials in contemporary nonlinear methods for the behavioral*. pp 26–94

Westaby JD, Pfaff DL, Redding N (2014) Psychology and social networks: a dynamic network theory perspective. Am Psychol 69:269–284. https://doi.org/10.1037/a0036106

Wiener N (1948) Cybernetics. Scientific American 179:14–19

Yerkes RM, Dodson JD (1908) The relation of strength of stimulus to rapidity of habit‑formation. J Comp Neurol Psychol 18:459–482. https://doi.org/10.1002/cne.920180503

Zanone PG, Kelso JA (1992) Evolution of behavioral attractors with learning: Nonequilibrium phase transitions. J Exp Psychol Hum Percept Perform 18:403–421. https://doi.org/10.1037/0096-1523.18.2.403